\documentstyle[11pt,newpasp,twoside]{article}
\def\edcomment#1{\iffalse\marginpar{\raggedright\sl#1\/}\else\relax\fi}
\reversemarginpar

\begin{document}
\title{Dynamics of line-driven disk winds}
 \author{Daniel Proga}
\affil{NASA Goddard Space Flight Center, 
Laboratory for High Energy Astrophysics, Code 662, Greenbelt, MD 20771}

\begin{abstract}
I review the main results from recent 2-D, time-dependent hydrodynamic models 
of radiation-driven winds from accretion disks in AGN. I also discuss 
the physical conditions needed for a disk wind to be shielded from 
the strong X-rays and to be  accelerated to hypersonic velocities. 
I conclude with a few remarks on 
winds in hot stars, low mass young stellar objects, cataclysmic variables, 
low mass X-ray binaries, and galactic black holes and future work.

\end{abstract}

\section{Introduction}

A key constraint on any model for the origin of AGN outflows is the 
ionization balance. On one hand we observe very high
luminosities in X-rays and the UV and 
on the other hand we observe spectral lines from moderately
and highly ionized species. One wonders then how the gas avoids full
photoionization and we see any spectral lines at all. Two
mechanisms have been proposed to resolve the so-called overionization  
problem: (i) the AGN outflows have filling factors less than one
and consist of dense clouds and (ii)  the filling factors
equal one but the outflows are shielded from the powerful radiation 
by some material located between the central engine and the outflow
(e.g., Krolik 1999).
Only limited citations will be possible due to space limitations;
my apologies in advance.

One plausible scenario for AGN outflows is
that a wind is driven from an accretion disk around a black hole. 
Radiation  pressure due to spectral lines is one of the forces that
have been suggested to accelerate outflows in AGNs.

Our understanding of how line-driving produces powerful high velocity
winds is based  on the studies of winds in hot stars
(e.g., Castor, Abbott, Klein 1975, hereafter CAK).  
The key element of the CAK model is that the momentum
is extracted most efficiently from the radiation field via line opacity.
CAK showed that the radiation force due to lines,
$F^{rad,l}$ can be stronger than the radiation force due to 
electron-scattering, $F^{rad,e}$ by up to several orders of magnitude 
(i.e., $F^{rad,l}/F^{rad,e} < M_{max} \approx 2000$). 
Thus even a star that radiates at around 0.05\% (i.e., $1/M_{max}$)
of its Eddington limit, $L_{E}$ can have a strong wind.

To apply line-driven stellar wind models to AGN 
we have to take into account at least two important differences:
(i) the difference in  geometry --
stellar winds are to a good approximation spherically symmetric,
whereas the wind in AGN likely arises from a 
disk and is  therefore axisymmetric; and (ii) the difference in 
the spectral energy distribution -- hot stars radiate mostly 
the UV (the UV luminosity, $L_{UV}$, accounts for most of the total 
luminosity, $L$), whereas AGN radiate strongly both in the UV and X-rays 
($L_{UV}$ and $L_X$ are comparable).
The latter difference has two important consequences on
UV line driving: not all AGN radiation contributes to driving,
and even worse, the X-rays that to do not contribute to UV line driving
can ionize the gas and reduce the number of transition that can scatter
the UV photons (in the case of a fully ionized gas, $F^{rad,l}=0$
and by definition $M_{max}=0$!).

The consequences of the difference in geometry have been recently studied 
using 2-D axisymmetric numerical hydrodynamical simulations
(e.g., Proga, Stone \& Drew 1998, hereafter PSD98). 
These simulations were focused on cataclysmic variables (CVs),
which as do hot stars, radiate mostly in the UV.
In particular, PSD98 explored the impact upon the mass-loss rate, $\dot{M}_w$ 
and outflow geometry caused by varying the system luminosity and 
the radiation field geometry. A striking outcome was that winds driven
from, and illuminated solely by, an accretion disk yield complex, unsteady
outflow. In this case, time-independent quantities can be determined only 
after averaging over several flow timescales. On the other hand, if  winds 
are illuminated by radiation mainly from the central object, then the disk 
yields steady outflow. PSD98 also found  that $\dot{M}_w$ is a strong 
function of the total luminosity, while the outflow geometry is determined by 
the geometry of  the radiation field. For high system luminosities, the disk 
mass-loss rate scales with the  luminosity in a way similar to stellar mass 
loss. As the system luminosity decreases below a critical value (about twice 
the effective Eddington limit, $L_{E}/M_{max}$) 
the mass-loss rate 
decreases quickly to zero. Matter is fed into the fast stream from within 
a few central object radii. In other words, the mass-loss 
rate per unit area decreases sharply with radius. 
The terminal velocity of the  stream 
is similar to that of the terminal velocity of a corresponding
spherical stellar wind, i.e., $v_\infty \sim {\rm a~few}~v_{esc}$,
where $v_{esc}$, is the escape velocity from the photosphere.
Thus the difference in geometry changes the wind geometry and time behavior
but has less effect on $\dot{M}_w$ and $v_{\infty}$.

Proga, Stone \& Kallman (2000, hereafter PSK) 
made another step in studying line-driven winds from accretion
disks. To assess how winds can be driven from a disk in the presence of very 
strong ionizing radiation (as in AGN) they adopted the approach from 
PSD98 with three major modifications: 1)  calculation of the parameters of 
the line force
based on the wind properties, 2) inclusion of optical depth effects on 
the continuum photons, and 3) inclusion of radiative heating and cooling of 
the gas. In next section I will review
results from PSK and from related calculations
by Proga \& Kallman (2001, PK hereafter).

\section{Hydrodynamical Simulations For AGN}

In PSK, 
we calculated a few disk wind models for the mass of the non-rotating black 
hole,  $M_{BH}=10^8~\rm M_{\odot}$. To determine the radiation field from 
the disk, we assumed the mass accretion rate 
$\dot{M}_a=1.8$~M$_{\odot}$~yr$^{-1}$. These system parameters yield the disk 
Eddington number, $L_D/L_{E}\equiv\Gamma_D=0.5$ and the disk inner radius, 
$r_\ast=8.8\times10^{13}$~cm. For the radiation field from 
the central engine, we assumed that one half to 
the central engine luminosity is radiated in the UV and the other
half in X-rays. 
The total central engine luminosity was assumed equal to
the  accretion disk luminosity.
To calculate the gas temperature, we assumed the temperature of 
the X-ray radiation, $T_{\rm X}=10$~KeV (see PSK for more details).

For a fixed disk atmosphere and central radiation source, the most important
parameter of the PSK model is the wind X-ray opacity
$\kappa_X$ that determines
the optical depth for the X-rays from the central object, $\tau_X$. 

Figure~1 presents the poloidal velocity  of the PSK wind model
for the X-ray opacity $\kappa_X=40~{\rm g^{-1}~cm^2}$ for the photoionization
parameter $\xi \equiv 4 \pi F_{\rm X}/n < 10^5$ and 
$\kappa_X=0.4~{\rm g^{-1}~cm^2}$ otherwise. Here $F_{\rm X}$ is the local
X-ray flux and $n$ is the number density of the gas.
The UV wind opacity, $\kappa_{UV}$ was assumed 
$0.4~{\rm g^{-1}~cm^2}$ for all $\xi$.
The arrows in Figure~1 show that the gas streamlines are perpendicular 
to the disk over some height that increases with radius. The streamlines then bend away from 
the central object and converge. The region where the flow is moving almost 
radially outward is associated with a high-velocity, high density stream. 
PSK's calculation follows (i) a  hot, low density flow with negative radial 
velocity in the polar region (ii) a dense, warm and fast equatorial outflow 
from the disk, (iii) a transitional zone in which the disk outflow is hot 
and struggles to escape the system.

\begin{figure}

\begin{minipage}[t]{2.8in}
\begin{picture}(120,70)
\put(500,-505){\includegraphics{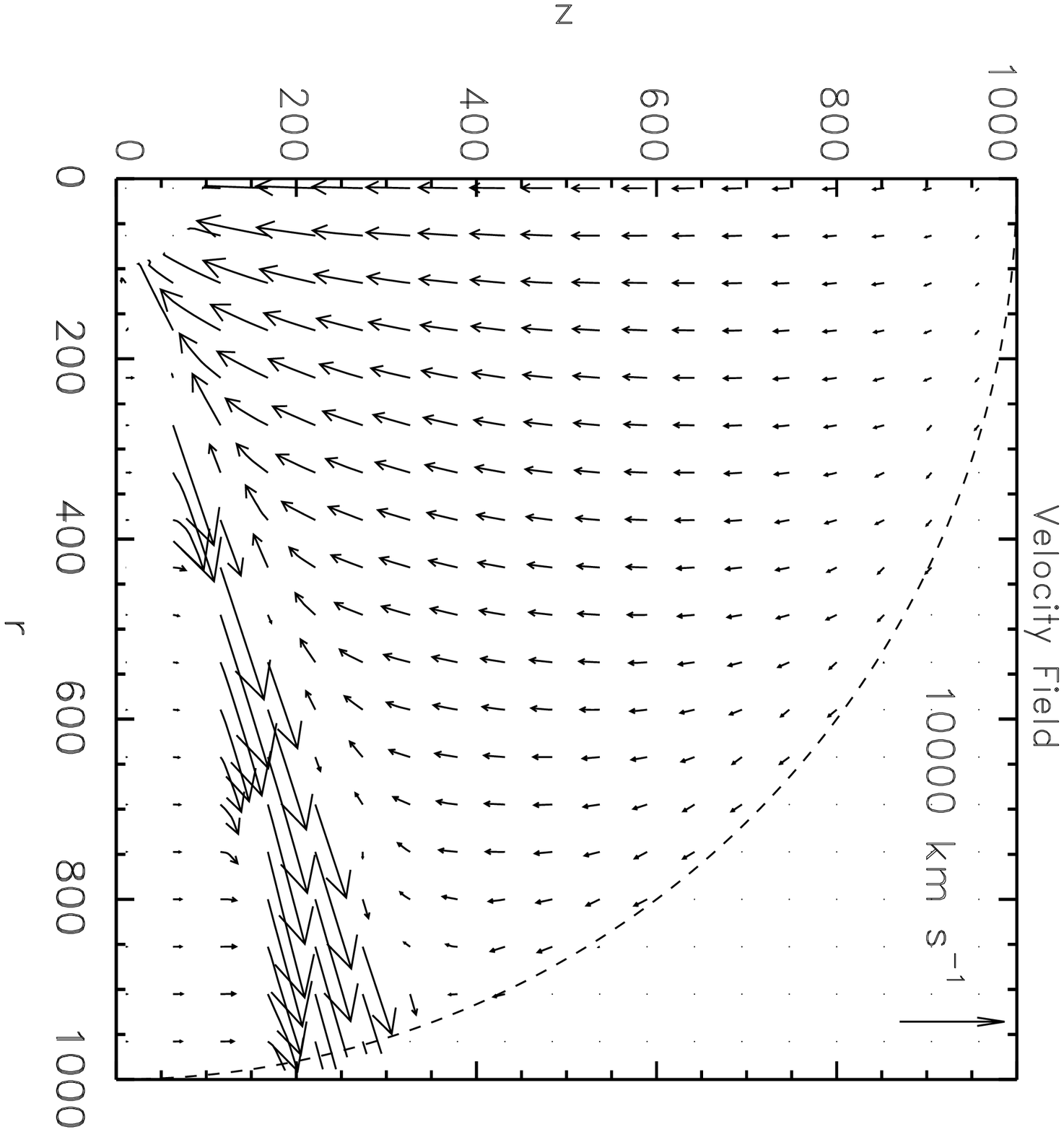}}
\end{picture}
\end{minipage} \hfill
\end{figure}
\begin{figure}
\vskip -1.2in
\begin{minipage}[b]{2.5in}
\caption{\small A map of the velocity field (the poloidal component only) 
for   
the line-driven  wind from a disk accreting on a $10^8~ M_{\odot}$ black hole
(PSK's Fig.~2). The rotation axis of the disk is along the left hand 
vertical frame, while the midplane of the disk is along the lower horizontal 
frame. The position on the figure is expressed in units of the disk inner 
radius (e.g., $100~r_\ast=8.8 \times 10^{15}$~cm).}
\end{minipage} 
\end{figure}

PSK found that the local disk radiation can launch a wind from the disk
despite  strong ionizing radiation from the central object.
The central radiation may overionize the supersonic portion of the flow
and severely reduce the wind velocity. To produce a fast disk wind
the X-ray opacity must be higher than the UV opacity by a factor $\ga$~100
for the photoionization parameter $\xi < 10^5$.
For lower relative X-ray opacity (i.e., $\kappa_X/\kappa_{UV}< 100$), 
the gas can be still launched from the disk by the  UV disk radiation 
but the gas velocity never exceeds the escape velocity. To radially
accelerate the gas lifted by the local disk radiation, the column density, $N_{\rm H}$
must be large enough to reduce the X-ray radiation but not too large to 
reduce the UV flux from the central engine. In other words, this requires 
$\tau_{\rm ~X}> 1$ and at the same time, $\tau_{UV} \ll 1$.

Our disk wind model can explain many aspects of AGN outflows. For example, 
the fast stream with the terminal velocity of  $\sim 15 000~{\rm km~s^{-1}}$ 
can be identified 
as a BAL region in QSOs. The stream density and column density  
are comparable to those observed in BAL QSOs. Because the stream is 
narrow, BALs should be seen only for a narrow range of inclination angles.
Synthetic line profiles calculated based of this model
confirm that the model can explain the observations (Proga 2001, 
in preparation). In particular, the line profiles
strongly vary with the inclination angle. The fast stream can produces
a strong, broad and blue-shifted absorption component, and even
multi-trough structure. Additionally, our model illustrates that  
terminal velocity of line-driven winds does not have to been coupled to 
$v_{esc}$,
it can be much smaller as in the transient zone in which the disk outflow
is overionized and loses support from the line force.  It is possible
that this slow inner flow, launched and driven by radiation, can 
explain warm absorbers and outflows in the Seyfert~I galaxies.

Let us return to the problem of launching a wind in the presence of X-rays:
when can gas be lifted from the disk by the local UV radiation
instead of being heated and overionized by the X-ray radiation?
This question can be qualitatively answered
in terms of the CAK model developed to study stars; results from PSD98 
for disks showed that this approach is reasonable.
For a given $L_E$, the stronger the UV radiation from the disk,
the higher the wind mass-loss rate and subsequently the wind density.
If we assume fixed X-ray radiation and no X-ray attenuation, 
we will find that if the UV radiation is strong enough, it can launch a wind
of so high density that the X-rays will be unable to ionize the wind.
The density of ionizing photons is simply  too low compared to the gas 
density, in other words the photoionization parameter will be low. 
The X-ray attenuation becomes essential when gas accelerates 
because the gas density decreases 
and eventually the X-ray may overionize the gas.

The line force is not always strong enough to push gas of sufficiently high 
density.
To examine in more detail what happens then, it is helpful to consider 
low mass 
X-ray binaries (LMXBs). LMXBs resemble AGN in a few respects, for example,
in their X-ray/bolometric luminosity ratios. Recent calculations  by PK 
found that in the case of LMXBs, the local disk radiation cannot launch a 
wind from the disk because of strong ionizing radiation from  the central 
object. Unphysically high X-ray opacities 
($\kappa_X\ge10^5~{\rm g^{-1}~cm^2}$) 
are required to shield the UV 
emitting disk and to allow the line force to drive a disk wind. 
However the same 
X-ray  radiation that inhibits  line driving heats the disk and can produce 
a hot bipolar wind or corona above the disk. PK's results 
are consistent with the UV observations of LMXB which show no obvious 
spectral features associated with strong and fast disk winds. 

\section{A Few Remarks}

An important aspect of studying AGN outflows is to understand why some 
AGN of the same type have outflows and some do not (e.g., QSOs with and 
without broad absorption lines, BALs), and why different types of AGN have 
outflows of different appearance (e.g., narrow UV absorption
lines  in the Seyfert~I galaxies and BALs in QSOs). The talks and posters
of this workshop have provided many wonderful examples of the variety of 
outflows in AGN. To address the problem of AGN outflows in general,
I will build upon theoretical results for line-driven winds I have discussed
so far. Fortunately, the 2-D hydrodynamical models 
can be understood fairly well using concepts from the original CAK model and 
the 2-D results can be approximated by analytic formulae.

\begin{figure}
\begin{picture}(100,120)
\put(120,-190){\includegraphics{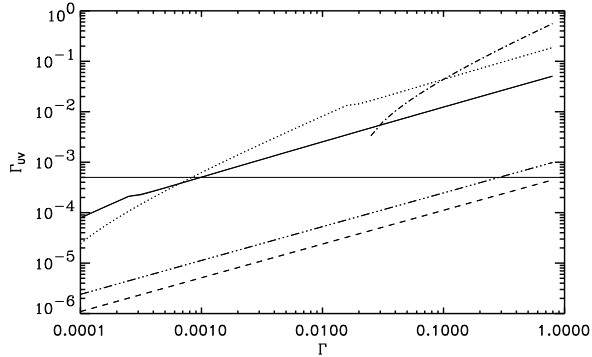}}
\end{picture}
\vskip 4mm
\caption{\small The UV Eddington number as a function of 
the total Eddington number for AGN, CVs, LMXBs, GBHs and 
FU Ori stars. See the text for the explanation of the lines.
}
\end{figure}

For example, it is possible to derive analytic formulae to estimate
the photoionization parameter at the base of a line-driven disk wind
(PK). PK 
compared results of line-driven disk wind models for accretion disks 
in LMXBs and AGN. They found that the key parameter determining 
the role of the line force is not merely the presence of the luminous UV zone 
in the disk and the presence of X-rays, but also the distance of 
this zone from the center. This result is not surprising because  
the closer the UV zone to the center, the higher the UV contribution
to the total luminosity. That in turn implies
a stronger  line force and subsequently 
a denser  disk wind launched by the line force. As I already mentioned, 
the density of the disk wind critically determines whether the wind will stay 
in a lower ionization state in the presence of X-ray radiation and be 
further accelerated by the line force to supersonic velocities.
Therefore in a general case, we ought to consider not
only the total luminosity but also the UV luminosity, $L_{UV}$
as well as the Eddington number of the UV emitting part of the 
disk, $\Gamma_{UV}$. 
The UV disk radiation is the one that is to drive a wind, the 
remaining disk radiation has either  little impact on the wind (i.e.,
the optical and infrared radiation of  cold disk)
or can reduce the line driving (i.e., X-rays  of
a hot disk). Simply put, the difference, $(1-\Gamma_{UV}) L$
is the luminosity that can 'damage' a wind emerging from the UV disk.

Figure~2 shows the UV Eddington number, $\Gamma_{UV}$ as a function of 
the total Eddington number, $\Gamma$ for various accreting objects: massive black hole  
($M_{BH}=10^8~\rm M_{\odot}$,  the dotted line), stellar black hole
($M_{BH}=10~\rm M_{\odot}$,  the triplet dot-dashed line),
neutron star ($M_{NS}=1.4~\rm M_{\odot}$ the dashed line),  
white dwarf ($M_{WD}=0.6~\rm M_{\odot}$ the solid line), and
low mass young stellar object ($M_{YSO}=0.2~\rm M_{\odot}$ 
the dot-dashed line).
Note that for fixed mass and radius of the accreting object,
$\Gamma$ is proportional to the mass accretion rate.
For simplicity, I estimate $L_{UV}$ by integrating
the disk intensity over the disk surface of the effective temperature
between 8,000~K and 50,000~K. Detailed photoionization calculations 
are needed to determine what is a contribution to the line force 
from radiation at the temperatures beyond this range.
To calculate the disk intensity and temperature
I used the standard  Shakura \& Sunyaev disk model 
(Shakura \& Sunyaev 1973). 
I define the UV Eddington number as the ratio between $L_{UV}$
and $L_{E}$. The solid horizontal line marks the UV Eddington number
above which line driving can drive a wind if there are  no X-rays in a system, 
$\Gamma_{UV} (M_{max}+1) < 1$. 
Inclusion of X-rays will move this line up
because the X-rays will reduce $M_{max}$.

Figure~2 allows an easy identification of objects 
capable of driving winds by lines. They are accretion disks
around: massive black holes and white dwarfs (AGN and CVs with
$\Gamma \ga 0.001$)
and low mass young stellar object 
(FU Ori stars with $\Gamma \ga$ a few $\times$ 0.01 !!!).
The systems that have too low  $\Gamma_{UV}$ to drive wind 
are accretion disks around low mass compact objects (LMXB and galactic
black holes, GBH). My classification of objects is consistent
with the observations in the sense that the objects I identified
as capable of driving winds have been observed to have winds
whereas the objects with too low $\Gamma_{UV}$ do not exhibit
strong spectral features associated with winds. Additionally my conclusions
based on this simple figure are consistent with detailed numerical
calculations (those for FU Ori and GBH have not been published).
I do not claim here that UV driving can fully explain the observed outflows
but simply make a point that UV driving can drive some winds.

The lesson from the above exercise is that if we  repeat it for AGN
of different masses of the black hole and different luminosities
(subdivided to  the UV and X-rays) we can gain some insight
into the nature of  all AGN outflows.

Future work in hydrodynamical simulations of mass outflows can only
benefit from simple analyses such as this. In fact,
there are a few difficult problems that one encounters while
making theoretical studies. For example, it is hard
to make detailed photoionization calculations in connection with
multidimensional time dependent calculations. However such
photoionization calculations are crucial to confirm if
the UV and X-ray opacities are such that the corresponding
optical depths are as required. Future work on hydrodynamical models
should include modeling the disk internal structure while
modeling disk winds. An important limitation of the PSK approach
is that it is valid for the gas pressure dominated disk.
PSK did not  include in their calculations the whole UV disk 
because its inner part is likely radiation dominated. 
Nevertheless they found that the mass-loss rate 
is about 25\% of the assumed mass accretion rate. 
If we assume that the whole UV disk
is gas pressure dominated, the mass loss rate will
exceed the mass accretion rate by a factor of 10 or more.
Does this result mean that the line force can drive such a strong wind 
that it can
significantly change accretion on a black hole? If this is true
then it would be very interesting because the models for line-driven
disk winds in CVs predict $\dot{M}_w$ that is too low to completely explain 
the observations. 

\acknowledgments
The work presented in this paper was performed while I held
a NRC Research Associateship at NASA/GSFC.
I thank  J.E. Drew, T.R. Kallman, D. Kazanas, S.J. Kenyon, and J.M. Stone
for useful discussions. I am grateful to S.J.K.  for  
bringing to my attention outflows in FU Ori stars.

\end{document}